\documentclass{ws-procs975x65}
\usepackage{graphicx}
\usepackage{amssymb}
\usepackage{epstopdf}
\usepackage[dvips]{color}
\newcommand{\msbar}{\overline {\rm MS}}
\newcommand{\SF}{Schr\"odinger functional\ }

\begin{document}

\title{Study of the running coupling constant in 10-flavor QCD with the Schr\"odinger functional method}

\author{N.~Yamada$^{a,b}$,  M.~Hayakawa$^{c}$, K.-I.~Ishikawa$^{d}$, Y.~Osaki$^{d}$,
            S.~Takeda$^{e}$, and S.~Uno$^{c}$}

\address{
$^a$ KEK Theory Center, Institute of Particle and Nuclear Studies, High
Energy Accelerator Research Organization (KEK), Tsukuba 305-0801,
Japan\\
$^b$ School of High Energy Accelerator Science, The Graduate University
for Advanced Studies (Sokendai), Tsukuba 305-0801, Japan\\
$^c$ Department of Physics, Nagoya University, Nagoya 464-8602, Japan\\
$^d$ Department of Physics, Hiroshima University, Higashi-Hiroshima 739-8526, Japan\\
$^e$ School of Mathematics and Physics, College of Science and
Engineering, Kanazawa university, Kakuma-machi, Kanazawa, Ishikawa
920-1192, Japan}
\begin{abstract}
 \vspace*{-128mm}
\begin{flushright}
  \normalsize
 KEK-CP-233,\ HUPD-0909,\ KANAZAWA-10-02
 \end{flushright}
 \vspace*{118mm}\
The electroweak gauge symmetry is allowed to be spontaneously broken by
the strongly interacting vector-like gauge dynamics.
When the gauge coupling of a theory runs slowly in a wide range of
energy scale, the theory is a candidate for walking technicolor.
This may open up the possibility that the origin of all masses may be
traced back to the gauge theory.
We use the \SF method to see whether the gauge coupling of 10-flavor QCD
``walks'' or not.
Preliminary result is reported.
\end{abstract}


\bodymatter

\section{Introduction}

The main goal of Large Hadron Collider (LHC) is to confirm the Higgs
mechanism and to find particle contents and the physics law above the
electroweak scale.
So far many new physics models beyond the standard model have been
proposed.
Among them, Technicolor (TC)~\cite{Weinberg:1975gm} is one of the most
attractive candidates~\cite{Hill:2002ap} as it does not require
elementary scalar particles which cause, so-called, the fine-tuning
problem.
This model is basically a QCD-like, strongly interacting vector-like
gauge theory.
Therefore, lattice gauge theory provides the best way to study this
class of model\cite{Fleming:2008gy}, and the predictions can be as
precise as those for QCD, in principle.

The simple, QCD-like TC model, {\it i.e.} an SU(3) gauge theory with two
or three flavors of techniquarks, has been already ruled out by, for
instance, the S-parameter\cite{Peskin:1990zt} and the FCNC constraints.
However, it has been argued that, if the gauge coupling runs very slowly
(``walks'') in a wide range of energy scale before spontaneous chiral
symmetry breaking occurs, at least, the FCNC problem may
disappear\cite{Holdom:1981rm}.
Such TC models are called walking technicolor (WTC) and several explicit
candidates are discussed in semi-quantitative manner in
Ref.~\cite{Dietrich:2006cm}.
Since the dynamics in WTC might be completely different from that in QCD
and hence the use of the naive scaling in $N_{c}$ or $N_{f}$ to estimate
various quantities may not work, the $S$-parameter must be evaluated
from the first principles~\cite{Shintani:2008qe}.
Although really important quantity is the anomalous dimension of
$\bar\psi \psi$ operator, looking for theories showing the walking
behavior is a good starting point.
Recently many groups started quantitative studies using lattice
technique to answer the question what gauge theory shows walking
behavior.
In Ref.~\cite{Appelquist:2007hu}, the running couplings of 8- and
12-flavor QCD are studied on the lattice using the \SF (SF)
scheme\cite{Luscher:1992an}.
Their conclusion is that while 8-flavor QCD does not show walking
behavior 12-flavor QCD reaches an infrared fixed point (IRFP) at
$g_{\rm IR}^{2}\sim 5$.
In spite of the scheme-dependence of running and its value of IRFP, the
speculation inferred from Schwinger-Dyson
equation~\cite{Appelquist:1988yc} suggests that $g_{\rm IR}^{2}\sim 5$
is not large enough to trigger spontaneous chiral symmetry breaking.
Although 12-flavor QCD is still an attractive candidate and is open to
debate~\cite{Hasenfratz:2009ea}, we explore other $N_{f}$.
In the following, we report the preliminary results on the running
coupling in 10-flavor QCD.
Since the conference, statistics and analysis method are changed.
The following analysis is based on the increased statistics and a
slightly different analysis method.

\section{Perturbative analysis}
\label{sec:ptanalysis}

Before going into the simulation details, let us discuss some results from perturbative analysis.
In this work, we adopt the $\beta$ function defined by
\begin{eqnarray}
     \beta(g^2(L))
 &=& L\,\frac{\partial\,g^2(L)}{\partial L}
  = b_1\,g^4(L)+b_2\,g^6(L)+b_3\,g^8(L)+b_4\,g^{10}(L)+\cdots,
\end{eqnarray}
where $L$ denotes a length scale.
The first two coefficients are scheme-independent, and given by
\begin{eqnarray}
     b_1
 = \frac{2}{(4\pi)^2}\left[ 11 - \frac{2}{3}N_f\right],
&\ \ \ &
     b_2
 = \frac{2}{(4\pi)^4} \left[\, 102 - \frac{38}{3}N_f\,\right].
\label{eq:b1-b2}
\end{eqnarray}
Other higher order coefficients are scheme-dependent and are known only
in the limited schemes and orders.
In this section, we analyze the perturbative running in the four
different schemes/approximations:
i) two-loop (universal),
ii) three-loop in the $\msbar$ scheme,
iii) four-loop in the $\msbar$ scheme,
iv) three-loop in the \SF scheme.
The perturbative coefficients relevant to the following analysis are
\begin{eqnarray}
     b_3^{\msbar}
 &=& \frac{2}{(4\pi)^6}
     \left[\, \frac{2857}{ 2}
            - \frac{5033}{18}N_f
            + \frac{ 325}{54}N_f^2\,
     \right],
\label{eq:beta-3MS}\\
     b_4^{\msbar}
 &=& \frac{2}{(4\pi)^8}
     \left[\, 29243.0
            - 6946.30\,N_f
            + 405.089\,N_f^2\,
            + 1.49931\,N_f^3\
     \right],\\
     b_3^{\rm SF}
 &=& b_3^{\msbar} + \frac{b_2\,c_2^{\theta}        }{2\pi}
                  - \frac{b_1\,(c_3^{\theta}-{c_2^{\theta}}^2)}{8\pi^2},
\label{eq:b3SF}
\end{eqnarray}
where the coefficients $c_2^{\theta}$ and $c_3^{\theta}$ depend on the
spatial boundary condition of the SF used in calculations, {\it i.e}
$\theta$.
Those for $\theta=\pi/5$ and $c_{2}^{\theta}$ for $\theta=0$ are known
as
\begin{eqnarray}
     c_2^{\theta=\pi/5}
 &=& 1.25563+0.039863\times N_f,\\
     c_3^{\theta=\pi/5}
 &=& ({c_2^{\theta=\pi/5}})^2 + 1.197(10) + 0.140(6)\times N_f - 0.0330(2)\times N_f^2,\\
     c_2^{\theta=0}
 &=& 1.25563+0.022504\times N_f,
\end{eqnarray}
but $c_3^{\theta=0}$ is not.
Therefore, the case iv) is studied with $\theta$=$\pi/5$.
Notice that in our numerical simulation $\theta$=0 and thus, rigorously
speaking, the example iv) is not applied to our numerical result.

\begin{table}[tb]
\centering
\begin{tabular}{c|ccccccc}
 $N_f$ & 4 & 6 & 8 & 10 & 12 & 14 & 16\\
\hline
 2-loop universal &&&& 27.74 & 9.47 & 3.49 & 0.52\\
 \hline
 3-loop SF & 43.36 & 23.75 & 15.52 & 9.45 & 5.18 & 2.43 & 0.47\\
 \hline
 3-loop $\msbar$ && 159.92 & 18.40 & 9.60 & 5.46 & 2.70 & 0.50\\
 \hline
 4-loop $\msbar$ &&& 19.47 & 10.24 & 5.91 & 2.81 & 0.50\\
\end{tabular}\\[2ex]
\caption{}
\small{Tab.~1~The IRFP from perturbative analysis.}
\label{tab:p-IRFP}
\end{table}
The perturbative infrared fixed point (IRFP) are numerically solved and
summarized in Tab.~\ref{tab:p-IRFP}.
As seen from Tab.~\ref{tab:p-IRFP}, with $N_{f}\ge 8$ the fixed point
value is, to some extent, stable against the change of
schemes/approximations.
It is interesting that the IRFP of 12-flavor QCD in the SF scheme is
consistent with the non-perturbative calculation by
Ref.~\cite{Appelquist:2007hu}.
Now looking at the perturbative IRFP at $N_{f}=10$, it appears to be
stable at $g^{2}\sim 10$.
Furthermore, according to the analysis based on Schwinger-Dyson equation,
there is an argument that chiral symmetry breaking occurs at around
$g^{2}\sim 4\pi^{2}/(3\,C_{2}(R)) = \pi^2$~\cite{Appelquist:1988yc}.
In summary, the perturbative analysis suggests that 10-flavor QCD is the
most attractive candidate for WTC.

\section{Simulation parameters and setup}

We employ the \SF method~\cite{Luscher:1992an} to calculate the running
coupling constant.
Unimproved Wilson fermion action and the standard plaquette gauge action
without any boundary counter terms are used.
The parameter of the spatial boundary condition for fermions, $\theta$,
is set to zero.
The bare gauge coupling $\beta=6/g^{2}_{0}$ is explored in the range of
4.4--24.0.
In this analysis, we report the results obtained from $(L/a)^{4}$ =
$6^{4}$, $8^{4}$ and $12^{4}$ lattices.
The calculation on $16^{4}$ lattice is in progress.
The numerical simulation is carried out on several architectures
including GPGPU and PC cluster.
The standard HMC algorithm is used with some improvements in the solver
part like the mixed precision algorithm.
So far, we have accumulated 5,000 to 200,000 trajectories depending on
$(\beta,\ L/a)$.

Since the Wilson type fermion explicitly violates chiral symmetry, the
critical value of $\kappa$ has to be tuned to the massless limit.
We performed this tuning for every pair of $(\beta,\ L/a)$.
At around $\beta\sim$ 4.4, for massless fermions we encounter a
(probably first order) phase transition independently of $L/a$, where
the plaquette value suddenly jumps to a smaller value.
Since this bulk phase transition is inferred to be lattice artifact,
whenever this happens we discard the configurations.
Thus the position of the critical $\beta$ ($\sim 4.4$) sets the lower
limit on $\beta$ at which simulations make sense.

\section{Preliminary results}

\begin{figure}[bt]
\centering
\begin{tabular}{cc}
\includegraphics*[width=0.5 \textwidth,clip=true]
{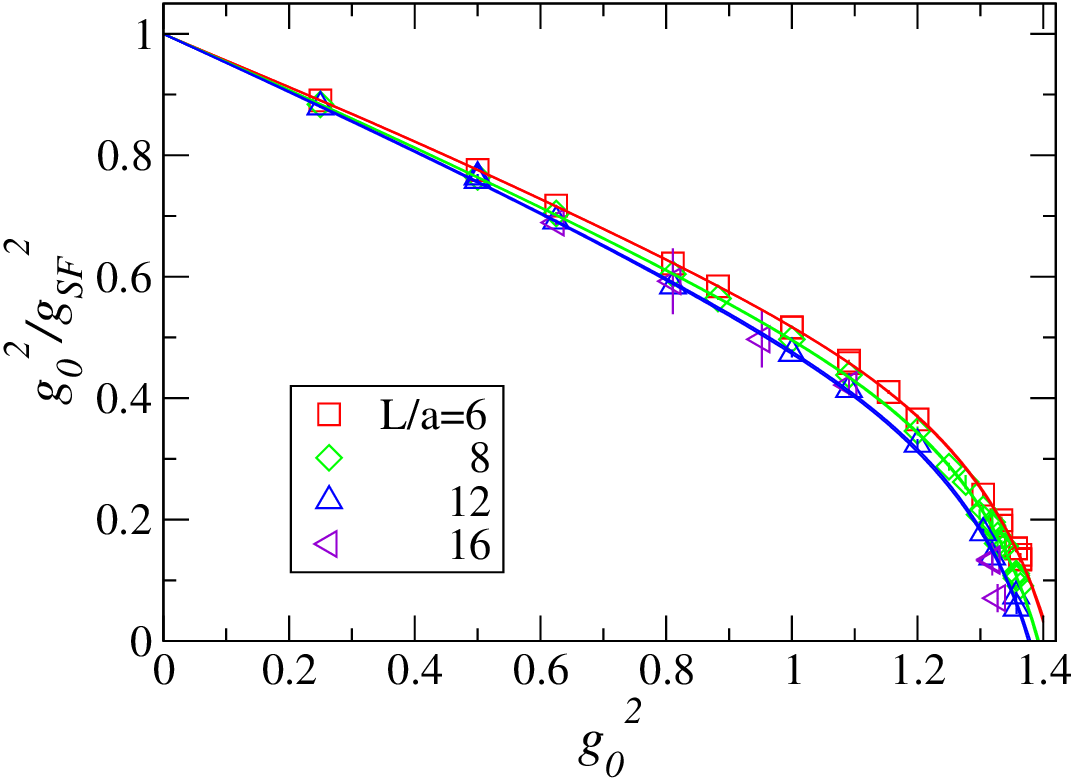} &
\includegraphics*[width=0.5 \textwidth,clip=true]
{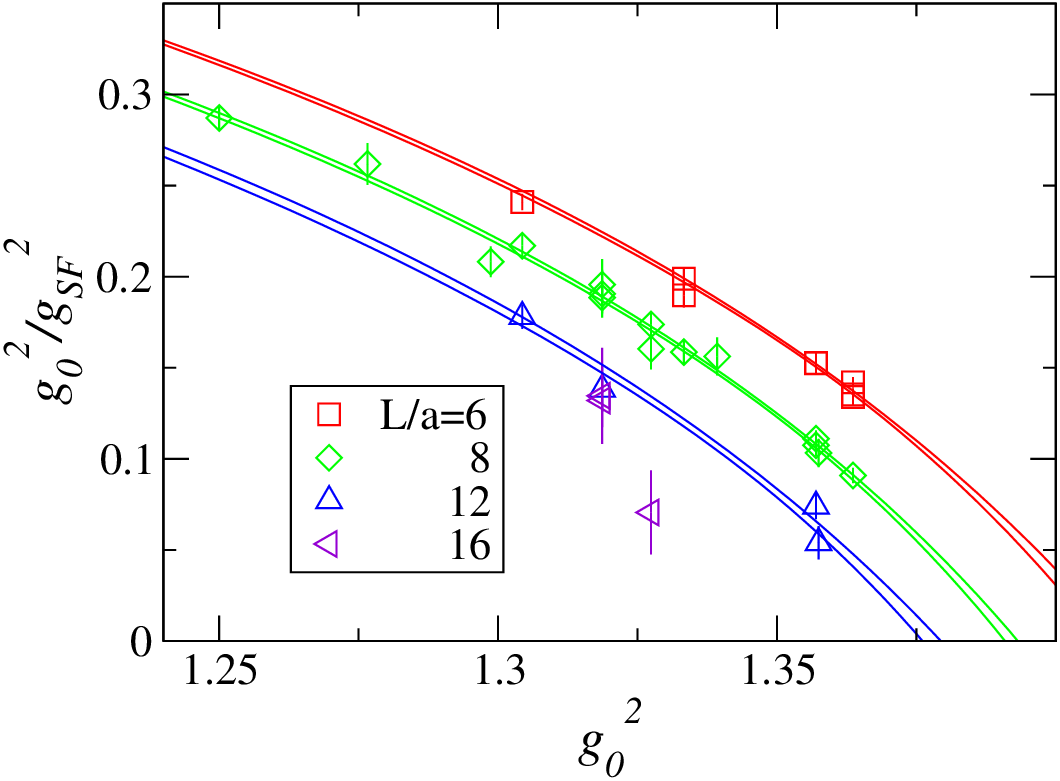}
\end{tabular}
\caption{$g_{0}^{2}$-dependence of $g_{0}^{2}/g_{\rm SF}^2(L)$ at
 $L/a=6,\ 8,\ 12,\ 16$. The right panel enlarges the region of $g_0^2\in [1.25,\ 1.40]$ of the
 left.}
\label{fig:main_result_1}
\end{figure}
Figure~\ref{fig:main_result_1} shows the \SF coupling calculated on the
lattices, where $g_0^2/g^{2}_{\rm SF}$ is plotted as a function of the
bare coupling.
The solid curve is the result (and statistical error) of the fit to
\begin{eqnarray}
     \frac{g_0^2}{{g_{\rm SF}^{\rm lat}}^2(g^{2}_{0}, L/a)}
 &=& \frac{1-a_{L/a,1}\,g_0^4}
          {1+p_{1,L/a}\times g_0^2+
           \sum_{n=2}^N a_{L/a,n} \times g_0^{2\,n}
          },
 \label{eq:fitfunc}
\end{eqnarray}
where $p_{1,L/a}$ is the $L/a$-dependent coefficient and is found, by
perturbative calculation, to be
\begin{eqnarray}
 p_{1,L/a} = \left\{
 \begin{array}{ll}
0.4477107831 & \mbox{ for } L/a=6\\
0.4624813408 & \mbox{ for } L/a=8\\
0.4756888260 & \mbox{ for } L/a=12 
 \end{array}
 \right..
\end{eqnarray}
We optimize the degree of polynomial $N$ in the denominator of
(\ref{eq:fitfunc}) by monitoring $\chi^{2}$/dof, and take $N=5$ for
$L/a$ = 6 and $N=4$ for $L/a$ = 8, 12.

Since we do not implement any $O(a)$ improvements, large scaling
violation is expected to exist.
One promising prescription to improve discretization errors has been
proposed in Ref.~\cite{Aoki:2009tf}.
Let us parameterize the lattice artifact in the step scaling by
\begin{eqnarray}
   \delta(u,s,L/a)
 = \frac{\displaystyle
         \frac{\Sigma_0(u,s,L/a)}
              {1+\delta^{(1)}(s,L/a)\,u}-\sigma(u,s)}
        {\sigma(u,s)}
  = \delta^{(2)}(s,L/a) u^{2}+ \cdots,
 \label{eq:Sig-sig}
\end{eqnarray}
where $u=g_{\rm SF}^{2}(L)$, $\sigma(u,s)=g_{\rm SF}^{2}(s L)$ and
$\Sigma_0(u,s,L/a)$ is ${g_{\rm SF}^{\rm lat}}^2(g_0^2,s L/a)$ at
$g_0^2$ satisfying ${g_{\rm SF}^{\rm lat}}^2(g_0^2,L/a)=u$.

The coefficient $\delta^{(1)}(s,L/a)$ in eq.~(\ref{eq:Sig-sig}) is
given by
\begin{eqnarray}
   \delta^{(1)}(s,L/a)
 = \Big( p_{1,sL/a}-b_{1}\ln(sL/a) \Big)
 - \Big( p_{1, L/a}-b_{1}\ln( L/a) \Big).
\end{eqnarray}
Dividing the lattice data $\Sigma_0(u,s,L/a)$ by the factor
$(1+\delta^{(1)}(s,L/a))$ improves the $O(u)$ discretization error and
hence $\delta(u,s,L/a)$ starts from $O(u^{2})$ as already indicated in
eq.~(\ref{eq:Sig-sig}).

$\sigma(u,s)$ has perturbative expansion,
\begin{eqnarray}
&&  \sigma(u,s) = u + s_0 u^2 + s_1 u^3 + s_2 u^4 + s_3 u^5  
       + \cdots,
       \label{eq:sigma}\\
&& s_0 = {b_1} \ln (s),\ \ \ \
      s_1= \ln (s) \left({b_1}^2 \ln (s)+{b_2}\right),\\
&& s_2 = \ln(s)\left( {b_1}^3 \ln^2(s)+\frac{5}{2} {b_1} {b_2} \ln(s)+{b_3} \right), \\
&& s_3
 = \ln(s)
    \left\{  {b_1}^4 \ln^3(s)+ \frac{13}{3} {b_1}^2 {b_2} \ln^2(s)
          + \ln (s) \left( 3{b_1}{b_3}+\frac{3}{2} {b_2}^2 \right)
          + {b_4}\right\},
\end{eqnarray}
where $b_i$'s are the coefficients of the $\beta$-function introduced in
sec.~\ref{sec:ptanalysis}.
Since we know the first two coefficients $b_{1}$ and $b_{2}$ and hence
$\sigma(u,s)$ can be numerically determined to $O(u^{3})$, with such
$\sigma$ the $O(u^{2})$ term in $\delta(u,s,L/a)$ is attributed to
discretization error.
The coefficient of $u^{2}$ term, $\delta^{(2)}(s,L/a)$, is then obtained
by fitting $\delta(u,s,L/a)$ to a quadratic function of $u$.
This fit must be done in the small-coupling region where the
perturbative series is reliable.
Since at this moment we have only a limited number of data points in
such a region, the fit range is forced to extend to $u\sim 2.5$.

\begin{figure}[bt]
\centering
\begin{tabular}{c}
\includegraphics*[width=0.6 \textwidth,clip=true]
{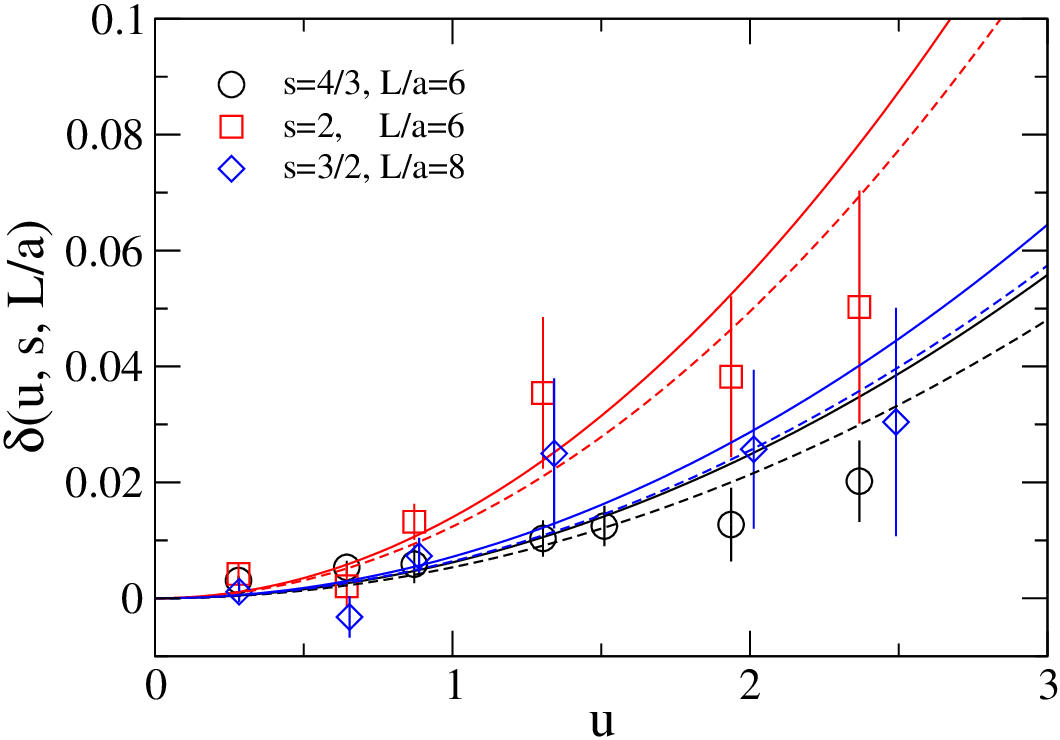}\\
\end{tabular}
\caption{$\delta$ as a function of $u$. $\delta^{(2)}(u,s,L/a)$ is
 determined as a coefficient of the quadratic term in $u$. (solid and
 dashed curves are obtained from different fit ranges.)}
\label{fig:delta_1_fit}
\end{figure}

Figure \ref{fig:delta_1_fit} shows the $u$ dependence of
$\delta(u,s,L/a)$ and the fit results.
The obtained coefficients are
\begin{eqnarray}
 \delta^{(2)}(s,L/a) =
  \left\{
  \begin{array}{ll}
   0.0062(9)  & \mbox{for } s=4/3,\ L/a=6\\
   0.0138(26) & \mbox{for } s=2,\   L/a=6\\
   0.0070(24) & \mbox{for } s=3/2,\ L/a=8
   \\
  \end{array}
  \right.,
  \label{eq:d2fit_1} \\
 \delta^{(2)}(s,L/a) =
  \left\{
  \begin{array}{ll}
   0.0053(8)  & \mbox{for } s=4/3,\ L/a=6\\
   0.0122(21) & \mbox{for } s=2,\   L/a=6\\
   0.0062(19) & \mbox{for } s=3/2,\ L/a=8
   \\
  \end{array}
  \right..
  \label{eq:d2fit_2}
\end{eqnarray}
In the fit, two different fit ranges,
$u\in [0,\ 2.02]$ and $[0,\ 2.50]$, are applied to examine the fit range
dependence of the result.
Eqs.~(\ref{eq:d2fit_1}) and eq.~(\ref{eq:d2fit_2}) correspond to the
former and the latter fit range, respectively.
Using $\delta^{(2)}(s,L/a)$ thus extracted, we define the improved
lattice data by
\begin{eqnarray}
   \Sigma^{\rm imp}(u,s,L/a)
 = \frac{\Sigma_0(u,s,L/a)}
        {1+\delta^{(1)}(s,L/a)\,u+\delta^{(2)}(s,L/a)\,u^{2}}. 
\label{eq:Sig_imp}
\end{eqnarray}

\begin{figure}[bt]
\centering
\begin{tabular}{ccc}
\includegraphics*[width=0.33 \textwidth,clip=true]
{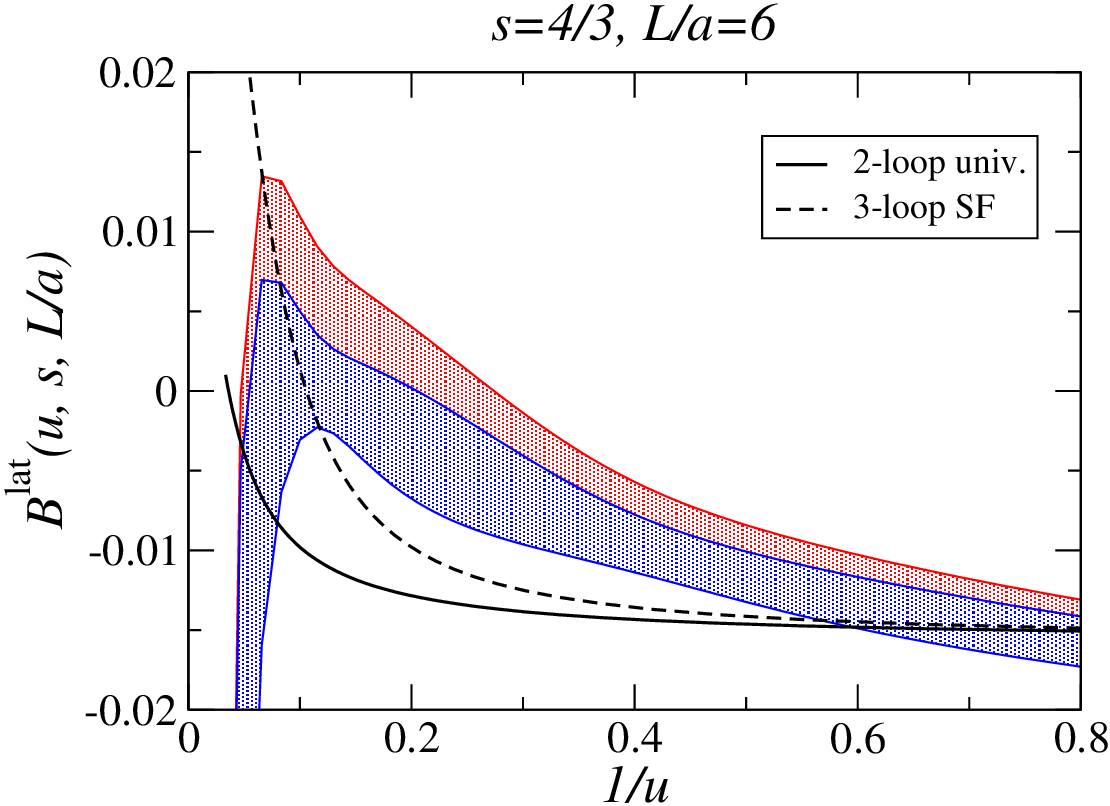}~
\includegraphics*[width=0.33 \textwidth,clip=true]
{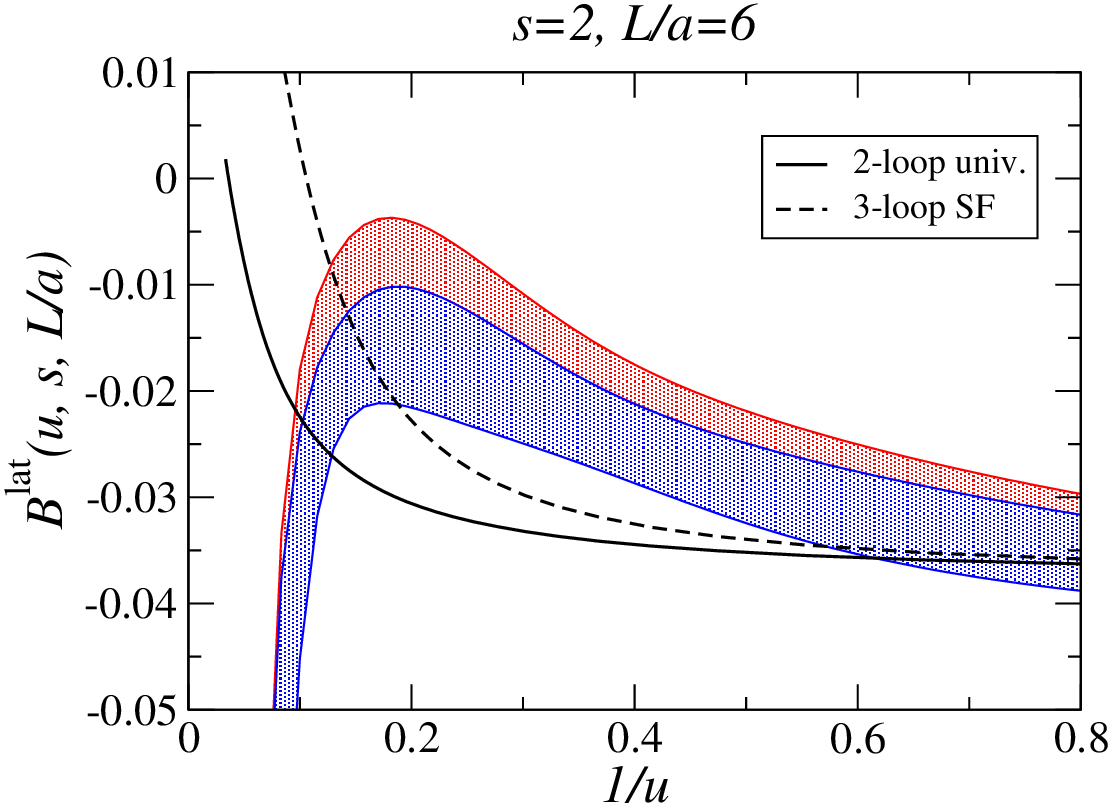}~
\includegraphics*[width=0.33 \textwidth,clip=true]
{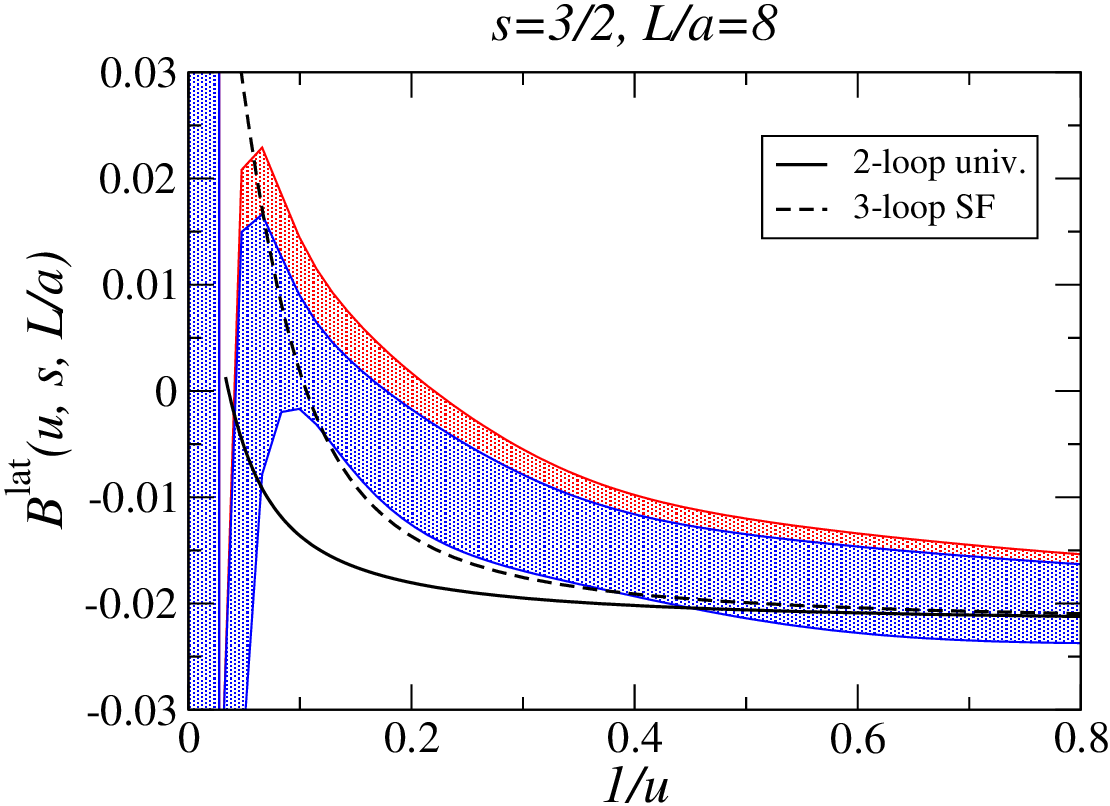}
\end{tabular}
\caption{Discrete beta functions for ($s,\ L/a$)=(4/3, 6), (2, 6) and
 (3/2, 8) from left to right. Two colored bands are the results with
 statistical error, obtained by the two-loop improvement with
 eq.~(\ref{eq:d2fit_1}) (red) and eq.~(\ref{eq:d2fit_2}) (blue),
 respectively. The black solid (dashed) curve is the corresponding
 discrete beta function obtained by integrating perturbative two-loop
 universal (three-loop SF scheme) beta function.}
\label{fig:dbf}
\end{figure}

To see the running in detail, we introduce the discrete beta
function~\cite{Shamir:2008pb},
\begin{eqnarray}
   B^{\rm lat}(u,s,L/a)
 = \frac{1}{\Sigma^{\rm imp}(u,s,L/a)} - \frac{1}{u}.
\end{eqnarray}
Figure~\ref{fig:dbf} shows the $1/u$ dependence of
$B^{\rm lat}(u,s,L/a)$.
As usual $\beta$ function, large negative value of $B^{\rm lat}$ means
rapid increase with length scale of the coupling, and $B^{\rm lat}$
flipping the sign indicate the existence of IRFP.
As seen from the figure, in either pairs of $(s,\ L/a)$ the discrete
beta function approaches to zero from below when $1/u$ decreases from
0.8 to 0.2, and this happens independently of the choice of
$\delta^{(2)}$.
This means that at around $u$=1.25 the running starts to slow down,
{\it i.e.} ``walk''.
In order to confirm that this behavior persists even in the continuum
limit, simulations on a larger lattice is necessary.

\section{Summary}

The running coupling constant of 10-flavor QCD is studied. 
The perturbative analysis suggests that this theory is extremely interesting.
The preliminary result obtained without continuum limit seems to
indicate the walking behavior.
In order to draw definite conclusions, we clearly need larger lattices
to take the continuum limit.
Such calculations are in progress.

\vspace{3ex}\
A part of numerical simulations is performed
on Hitachi SR11000 and the IBM System Blue Gene Solution at High Energy
Accelerator Research Organization (KEK) under a support of its Large
Scale Simulation Program (No. 09-05),
on GCOE (Quest for Fundamental Principles in the Universe) cluster
system at Nagoya University and
on the INSAM (Institute for Numerical Simulations and Applied
Mathematics) GPU cluster at Hiroshima University.
This work is supported in part by the Grant-in-Aid for Scientific Research
of the Japanese Ministry of Education, Culture, Sports, Science and Technology
(Nos. 20105001, 20105002, 20105005, 21684013, 20540261 and 20740139), and
by US DOE grant \#DE-FG02-92ER40699.

\end{document}